# Virtualization Technologies and Cloud Security: advantages, issues, and perspectives

Roberto Di Pietro[1], Flavio Lombardi[2]

[1] College of Science and Engineering, Hamad Bin Khalifa University - Doha, Qatar
rdipietro@hbku.edu.qa
[2] Istituto per le Applicazioni del Calcolo, Consiglio Nazionale delle Ricerche - Rome, Italy
flavio.lombardi@cnr.it

**Abstract.** Virtualization technologies allow multiple tenants to share physical resources with a degree of security and isolation that cannot be guaranteed by mere containerization. Further, virtualization allows protected transparent introspection of Virtual Machine activity and content, thus supporting additional control and monitoring. These features provide an explanation, although partial, of why virtualization has been an enabler for the flourishing of cloud services. Nevertheless, security and privacy issues are still present in virtualization technology and hence in Cloud platforms. As an example, even hardware virtualization protection/isolation is far from being perfect and uncircumventable, as recently discovered vulnerabilities show. The objective of this paper is to shed light on current virtualization technology and its evolution from the point of view of security, having as an objective its applications to the Cloud setting.

## 1 INTRODUCTION

The advances in virtualization technology of the past decade have rendered the Cloud approach feasible and convenient. Nevertheless, the main limitation of virtual machines is that they were born as a means to easily migrate from physically deployed services to more compact and manageable images. In fact, each and every VM runs its own full operating system together with the various libraries required by the application (see Fig. 1) [36]. Such an approach multiplicates the usage of RAM, CPU, and storage with respect to simply hosting multiple services as separate processes on a single piece of bare metal.

Containerization technology is intended to replace hypervisor and VMs, and deploys each application in its own process-like environment running on the physical machine on a single operating system [43]. Containers can be provisioned (and deprovisioned) in a few seconds and make a more efficient usage of resources, achieving a much higher application density (orders of magnitude [38]) than virtualization. This renders containers much more convenient than virtual machines.

Nevertheless, as we will show along this paper, virtualization is not on a dead path. In fact, virtual machines provide additional security mechanisms and isolation benefits in many application scenarios that are often worth the additional resource usage [28, 40].

A virtualization environment generally consists of three core components: an hypervisor or Virtual Machine Manager (also VMM in the following), management tools, and Virtual Machines (VMs). In particular, the infrastructure-as-a-service (IaaS) Cloud layer directly leverages and exposes powerful virtualization technologies and resources to a remote user [3]. Nevertheless, virtualization technologies also introduce additional security concerns. The size of the attack surface for the virtualization approach is directly proportional to the amount of emulated physical resource or functionality that must be provided in software. As regards containers, they can leverage all services offered by the host OS. The issue here is to enforce effective security and isolation among processes. This is actually more difficult to do, since OSes have not been designed with this in mind. Further, the partitioning/virtualization modes and ISAs [1] of recent CPU and GPU cannot be used by containers, as they are inherently part of virtualization and introduce the actual performance penalties of traditional VMs. Unikernels can be considered an alternative to both containerization and virtualization. Their main strength lies in being more lightweight than virtualization and better isolated than containers. Nevertheless, unikernels introduce further issues such as manageability, monitoring, and reliability.

In this paper, we survey various aspects of virtualization, analyze their impact on security, and discuss future perspectives. In particular, we provide technology background for most widespread virtualization tools in order to highlight features, advantages, and potential security flaws, with a focus on their application to Cloud. Further, discussions and comparisons with containerization and unikernel approaches are introduced.

The sequel of this paper is organized as follows: a technology background is provided in Section 2; most relevant virtualization security issues are introduced in Section 3; virtualization-based security approaches are presented in Section 4; novel enclave technology is discussed in Section 5; virtualization-based use cases, together with some future research trends, are presented in Section 6; and, finally, conclusions and hints for future work are given in Section 7.

## 2  Technology Background

Various different virtualization technologies are currently deployed in the Cloud, mostly for x86_64 architectures (e.g., Xen, KVM, VMware, VirtualBox, and HyperV). Most relevant details on virtualization frameworks and on supporting hardware (CPU/GPU) features are given and discussed in the following sections.

### 2.1  Virtualization Frameworks

The essential characteristics of the most widespread virtualization environments are summarized in Table 1. It is worth noting that all present hypervisors support full virtualization (also hardware-assisted virtualization in the following), as it offers relevant performance and isolation benefits. In fact, hardware virtualization allows the CPU to detect and possibly block unauthorized or malicious access to virtual resources. Nevertheless, no virtualization framework is immune to bugs. The virtualization platform can be an additional attack surface.

---

[1] Instruction Set Architecture(s)

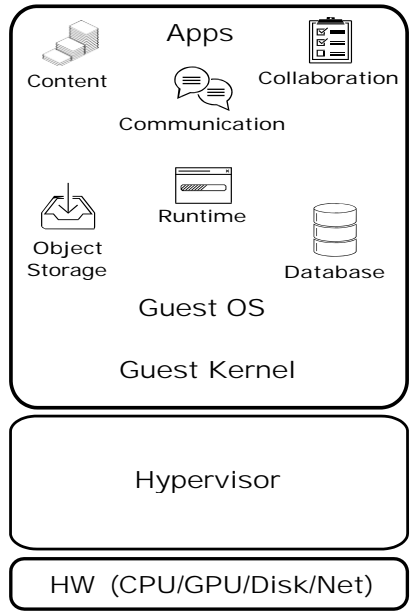

**Fig. 1.** Cloud layers and Virtualization

### 2.2  CPU Virtualization

The introduction of virtualization-enabling extensions in Intel and AMD CPUs dates back to 2005 [1, 25]. VT-x and AMD-V were developed to add an additional more privileged execution ring where an hypervisor or Virtual Machine Manager (VMM) could supervise actual access to physical resources from less privileged execution rings, as depicted in Fig. 2.

**Table 1.** CPU-related Virtualization Features

| X86_64 Hypervisor | open source | hypervisor type | supported extension(s) |
|---|---|---|---|
| Xen | Y | Native | VT-x, AMD-V, EPT, RVI, VT-d, AMD-Vi |
| KVM | Y | Hosted | VT-x, AMD-V, EPT, RVI, VT-d, AMD-Vi |
| VMWare ESX | N | Native | VT-x, AMD-V, EPT, RVI, VT-d, AMD-Vi |
| Hyper-V | N | Native | VT-x, AMD-V, EPT, RVI, VT-d, AMD-Vi |
| VirtualBox | Y | Hosted | VT-x, AMD-V |

CPUs are required to support some advanced extensions in order to allow the hypervisor to leverage them, as can be seen in Table 1. More in detail:

- **Intel VT-x AMD-V**: These two CPU capability sets are the basic ingredients of hardware-supported virtualization. They introduce *Ring -1* allowing a guest virtual machine to run its kernel at standard privilege level (i.e., Ring 0);
- **Intel EPT, AMD RVI**: Rapid Virtualization Indexing and Extended Page Tables, i.e., the Support for Second Level Address Translation (SLAT) that can significantly improve performance;
- **Intel VT-d, AMD-Vi**: These CPU capabilities (directed I/O) allow faster I/O resource virtualization.

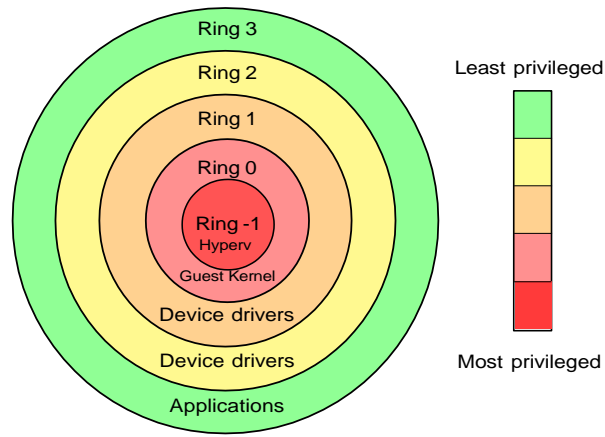

**Fig. 2.** Execution Rings for the x86_64 Architecture. See also [19]

### 2.3 GPU Virtualization

The virtualization paradigm also applies to Graphics Processing Units (GPUs). Virtual machines can be given mediated or full access to GPU computing and memory resources. This allows offering a GPU-based Cloud similar to what is already in place for CPU-based computing resource sharing. Hypervisor support for GPU virtualization features (see Table 2) is still somehow limited as relevant GPU technology is still reserved for high-end GPUs. In fact, GPU virtualization is usually implemented following one of these main approaches [24]:

- **time-sharing**: a single VM at a time is given direct access to the GPU. Time-slots are handled by the hypervisor;
- **passthrough**: the GPU is directly and permanently connected to a single VM that has direct access to it;

– **partitioned**: the GPU resources are split into smaller virtual GPUs, assigned to single VMs.

Once VMs have access to the GPU, the interaction between the guest and the real resource can be achieved in two different ways: backend virtualization or frontend virtualization [17]. Backend virtualization gives a direct connection between the VM and the GPU hardware. Frontend virtualization poses an intermediate layer between the guest and the hardware that has to leverage some kind of intermediate APIs to access the GPU. Some frontend virtualization examples are gVirt [57], vCUDA [54], GViM [22] and VOCL [60].

Table 2. GPU-related Virtualization Features

| X86_64 Hypervisor | open source | supported GPU Virtualization Technologies |
|---|---|---|
| Xen | Y | Intel GVT-g ,AMD MxGPU |
| KVM | Y | Intel GVT-g ,AMD MxGPU |
| VMWare ESX | N | Intel GVT-g ,AMD MxGPU |
| Hyper-V | N | - |
| Virtualbox | Y | - |

Particularly relevant here is AMD MxGPU technology [59], a partitioning strategy allowing users to have an equal share of the GPU. This hardware-based virtualization solution helps guaranteeing some isolation among different workloads and users.

Intel GVT-g [57] is a full GPU virtualization solution with mediated passthrough (VFIO[2] mediated device framework based). A virtual GPU instance is maintained for each VM, with part of performance critical resources directly assigned. The capability of running native graphics driver inside a VM, without hypervisor intervention in performance critical paths, achieves a good balance among performance, feature, and sharing capability.

As GPUs are mainly used for computation tasks, security concerns about GPU virtualization are mainly focused on data leakage [16]. This can occur either by directly access data owned by the victim and stored within the GPU memory or by exploiting side channels. In [42], Christin et al. have depicted two adversary models:

– **serial adversary**: this attacker has access to the victim's GPU or GPU memory, before or after the victim. Hence, it can seek for traces/data previously left in different GPU memories;
– **parallel adversary**: this attacker has contemporary access to the victim's GPU.

## 3 Virtualization Security Issues

Virtualization technologies underlying Cloud computing infrastructure themselves constitute vulnerable surface. In a Cloud scenario, we can observe the following major security challenges [36]:

---

[2] Virtual Function I/O

- **privileged user access**: access to sensitive data in the Cloud has to be restricted to a subset of trusted users (to mitigate the risk of abuse of high privilege roles);
- **lack of data/computation isolation**: one instance of customer data has to be fully isolated from data belonging to other customers;
- **reliability/availability**: the Cloud provider has to setup an effective replication and recovery mechanism to restore services, should a security issue occur;

Virtualization potentially widens Cloud computing attack vectors such as:

- **hypervisor**: the hypervisor is the software element sitting in between the host and guests to allow mediated access to physical resources. This layer should be transparent to a non-privileged user running into the guest. Unfortunately, its presence cannot be fully hidden [47]. As such, an attacker can exploit hypervisor vulnerabilities to gain access to both the host system and other guests. Hypervisors also provide emulation capabilities for missing hardware elements. However, this is a potential attack surface, as demonstrated by Ray [48] and Jason [26];
- **pivoting**: users can often login into specific services hosted by a VM. Once inside, the attacker could also exit the virtual machine she accessed, to damage the underlying physical system and/or sibling VMs.
- **migration**: virtual machines can be moved over different hosts for load balancing or disaster recovery. This "migration" is performed by copying the VM image over the network. An attacker can potentially eavesdrop data and perform a man in the middle attack if the channel is not encrypted.
- **resource allocation**: virtual machines are usually executed on-demand at run-time, thus making the resource allocation and management process as dynamic as possible. Resource sharing can thwart the security of the host system as well as of its virtual machines. In fact, negligence in cleaning resources before releasing them to others can lead to severe data leakage. As an example, data written by a VM into volatile or persistent storage can be accessed by others who have access to the same elements [51];

The above attacks show how virtual machines and the physical machines hosting them can be thwart by attackers targeting the host or just the virtual machine. Some mitigating approaches can be as follows:

- **host side**: vulnerabilities in the implementation of the hypervisor can somewhat be mitigated by frequently updating the hypervisor to reduce 0-days vulnerability window;
- **network monitoring**: monitoring and analyzing internal communications between sibling guests can help; nevertheless, malicious network behavior is difficult to detect by means of traditional intrusion detection systems and intrusion prevention systems;
- **encryption**: to mitigate such migration attacks, encryption of the data in transit can be used; nevertheless, this proves quite demanding on performance, and consequently on costs.
- **on allocation**: this attack can be dealt with by carefully deleting/cleaning resources either persistent or volatile that have been previously assigned to other VMs;

### 3.1 Co-Location issues

Co-location of virtual machines by different tenants on the same physical host is particularly frequent in Cloud computing. Virtual resources assigned to a tenant might get hacked by other virtual resources assigned to different tenants that are co-located within the same physical machine. Co-location can lead to different issues as follows:

- **information leakage**: by reusing the same physical hardware to allocate virtual resources, tenants might be able to exploit forensic tools to recover sensitive data from previous tenants;
- **performance degradation**: malicious tenants co-located in the same physical host might be able to make an uneven/widely varying use of computational power with high cpu-intensive co-located virtual machines with the final goal of degrading victim's performances;
- **service disruption**: malicious tenants sharing physical resources with their victim might be able to lead the hardware to unexpected behaviors thus causing a service disruption against the victim.

A large number of research results have highlighted the actual existence of co-location vulnerabilities [49, 62]. Such papers show that completely preventing tenants from sharing the same physical resources is practically unfeasible (due to rising costs). A viable solution [3] might be an attribute-based approach where tenants can express constraints over both virtual and physical resource allocation. Tenants would be able to indicate an high data sensitivity, thus requesting to avoid co-location. In this way, co-location will not be allowed for virtual resources working on high sensitive information thus lowering the chance of data leakage. As a consequence, virtual resource cost would be increased. This could be an acceptable trade-off in most sensitive scenarios.

### 3.2 Randomness and Virtualization

Cloud providers usually deploy identical VM clones when needed to satisfy request load. As such, it can happen that very similar (oftentimes the very same) images are (re)used for different tenants. As a consequence, the internal random pool for clone VMs is most probably the same/very similar for different VMs [20]. An adversary might exploit this weakness and try to guess the value of VM cryptographic keys [50]. In order to address such issue, the Cloud or Service providers should try to increase the number of events fed to the entropy pool of VM operating systems as soon as they are deployed, so as to provide an adequate level of security.

### 3.3 Container Security

The need for cost savings and shorter development cycles enabled the succes of containers in the Cloud. Containers are lighter than virtual machines and provide near-native performance. Docker[18] is the current market leader, providing a fully-featured packaging tool. Nevertheless, as introduced above, Containers provide much less isolation to applications, as such mechanisms are not based on hardware features but on process isolation approaches. Among other interesting works, Martin et al. [10] discuss Docker security real-world implications define an adversary model and describe several vulnerabilities affecting current Docker usage. The very same authors [41] detail Docker vulnerabilities and identify several vulnerabilities present by design or introduced by some original use-cases. Albeit some practical countermeasures are proposed, it is clear the containerization approach cannot guarantee an adequate level of security and protection in many multi-tenant scenarios.

### 3.4 Unikernel Security

The container limitation in providing actual isolation can be addressed by Unikernels, leveraging hardware virtualization to provide a potentially better alternative to containers (at least from the security point of view). Unikernels are specialized lightweight virtual machines (VMs) that squeeze the guest operating system and userspace layers together into one single VM layer [39]. This provides a smaller footprint, and a minimal attack surface. However, managing the privileges of thousands of unikernels is often difficult and error prone. An interesting approach is proposed in VirtusCap [53], a multi-layer access control architecture and mechanism leveraging unikernels. VirtusCap limits privileges of unikernels using the Principle of Least Privilege. This allowd creating unikernels that have only the privileges they need to accomplish their task.

### 3.5 Virtualization and Spectre/Meltdown

Spectre [31] and Meltdown [35] are recently discovered CPU vulnerabilities stemming from hardware-implemented performance optimizations aimed at reducing CPU-memory access latencies. Spectre leverages the fact that the speculative execution resulting from a branch misprediction can leave observable side effects that may reveal private data to attackers. In fact, when the memory access pattern depends on private data, the resulting state of the data cache constitutes a side channel an attacker can leverage to extract information about the private data.

Meltdown allows a userspace process to read all memory, even beyond its access scope. Like Spectre, the problem lies with speculative machine code execution that allows cache-timing attacks to leak data from all the memory.

Both Spectre and Meltdown are serious security vulnerabilities, in particular since they have been proven to even bypass CPU isolation features guaranteed by hardware-assisted virtualization. The reason why is that they are tied to hard-coded CPU optimizations that involve reusing (i.e. not deleting) cached values even though they belong to different (even security) contexts. Nevertheless, Containers and Unikernels are also vulnerable. As such, mitigating such hardware/firmware bugs is mandatory for any kind of co-location and multi-tenancy of the same physical CPU.

## 4 Virtualization Benefits for Security

Virtualization technologies also constitute a privileged point of view for observing and tracing VM activity. This can be used to collect useful data, analyze them, and act accordingly.

### 4.1 Virtual Machine Monitoring

A core set of requirements that a security monitoring system for the Cloud should meet can be summarized as follows [36]:

- **effectiveness**: the system should be able to detect attacks and integrity violations.
- **accuracy**: the system should be able to avoid false-positives, i.e., mistakenly detecting malware attacks where authorized activities are taking place.
- **transparency**: the system should minimize detectability from inside guests, i.e., potential intruders should not be able to detect the presence of the monitoring system.
- **robustness**: the host system, Cloud infrastructure and the sibling VMs should be protected from attacks proceeding from a compromised guest and it should not be possible to disable or alter the monitoring system itself.
- **reactivity**: the system should either be able to take action against both the attempt and the compromised guest, or notify other security-management components.
- **accountability**: the system should not interfere with Cloud and Cloud application actions, but collect data and snapshots to enforce accountability policies.

Nevertheless, satisfying these requirements is quite difficult, as there is a clear trade-off between transparency and reactivity. Possible mitigation approaches include:

- **hiding reaction**: i.e., leveraging regular guest maintenance actions as a reaction. E.g., halting the guest, restarting a fresh image, migrating the VM instance.
- **delaying reaction**: snapshotting the current status and delaying performing reactive activity. Nevertheless, the adversary might be able to perform further activity before being stopped.

In fact, a viable approach to achieve integrity protection is to continuously monitor key components that would most probably be targeted by attacks. We have shown (see also [36]) that by either actively or passively monitoring kernel or middleware components, it is actually possible to detect modifications to kernel data and code, thus guaranteeing that kernel and middleware integrity have not been compromised. A fully asynchronous monitoring system can be a viable solution [15] to provide protection and advanced transparent introspection capabilities to an hypervisor, as detailed in the following.

### 4.2 Semantic Introspection and Modeling VM Behavior

Monitoring key Cloud components that would be targeted or affected by attacks is vital in order to protect the VMs and the Cloud infrastructure [2]. By either actively or passively monitoring key VM components any possible modification to VM data and code can be traced and recorded.

In fact, virtual machine introspection is a process that allows observing the state of a VM from outside of it. Syringe [7] is one example of a monitoring system making use of virtualization to observe and monitor guest kernel code integrity from a privileged

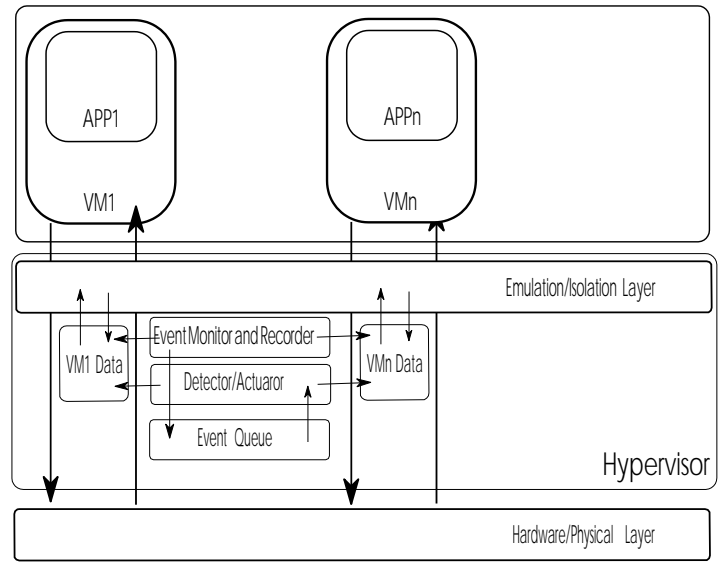

**Fig. 3.** Virtualization: Introspection Components

VM or from the VMM. However, it is quite simple for guest code to realize it is running inside a VM that can potentially be a honeypot VM [34].

The approach depicted in Fig. 3 is an example of advanced transparent passive tracing and recording of VM events from the hypervisor [36]. Any relevant event or status change is recorded by an event interceptor and it is then stored in a pool of recorder warnings where the collected information is asynchronously evaluated (evaluator) and, if needed, a reaction is triggered (act) according to a chosen policy.

An interesting VM-introspection-based approach is CloRExPa [15], providing various kinds of customizable resilience service solutions for Cloud guests, using execution path analysis. CloRExPa can trace, analyze and control live VM activity, and intervened code and data modifications, possibly due to either malicious attacks or software faults. Execution path analysis allows the VMM to trace the VM state and to prevent such a guest from reaching faulty states, leveraging scenario graphs.

This trend towards semantic introspection of VM activity is a very active field also as regards mobile devices in the Cloud [27]. This is the way to go for enabling control over possibly untrusted mobile Cloud nodes/applications. In fact, as will be detailed in the following for Bring Your Own Device (BYOD) untrusted devices, either they have to be banned altogether from the enterprise or enhanced semantics-aware introspection has to be put in place to prevent them from leaking sensitive information. Outside of the enterprise, semantic introspection allows legitimate users to regain control over their device internals. This approach will help detect and react to malware and to backdoors that are put in place even by trusted software or apps.

The main problem with introspection is that it requires knowing the internals and semantics of guest operating systems and running applications. This is especially difficult in case of closed-source OS and application such as in Windows and Mac environments. In fact, Windows OSes have always been the main target of malware that have exploited numerous bugs and vulnerabilities exposed by its implementations [37]. Recent trusted boot technology plus additional integrity checks have rendered the Windows OS less vulnerable to kernel-level rootkits [30]. Nevertheless, guest Windows Virtual Machines are becoming an increasingly interesting attack target. HyBIS [14] is the only example of introspection system protecting present Windows OS Guests from malware and rootkits.

### 4.3 Finer-Grained Security

Some other approaches are available that can enhance a general advanced protection system or be considered as a standalone solution.

As an example, Cloudvisor [61] is a transparent, backward-compatible approach protecting the privacy and integrity of cloud VMs. Cloudvisor separates the resource management from security protection in the virtualization layer. A small security monitor hidden under the VMM and leveraging nested virtualization [56] is used to protect the VMM and VMs. This approach is claimed of not affecting the security of users data inside the VMs.

In NestCloud [45] nested virtualization can be used in several usage models such as debugging and live migration. NestCloud is a three-level nested virtualization architecture minimizing the overhead caused by the additional level. NestCloud is a very effective approach for detailed introspection of VMs at the cost of increased latency and reduced performance.

Albeit not directly applied to cloud computing, Payer and Gross [46] presented an interesting work on virtualization for safe execution of applications based on software-based fault isolation and policy-based system call authorization. A running application is encapsulated in an additional layer of protection using dynamic binary translation in user-space. This virtualization layer dynamically recompiles the machine code and adds multiple dynamic security guards that verify the running code to protect and contain the application. The binary translation system implemented in [46] redirects all system calls to a policy-based system call authorization framework. This interposition framework validates every system call based on the given arguments and the location of the system call. Depending on the user-loadable policy and an extensible handler mechanism the framework decides whether a system call is allowed, rejected, or redirect to a specific user-space handler in the virtualization layer.

Also Lee et al. [32] discuss how new hardware architectural features for cloud servers can help protect the confidentiality and integrity of a cloud customer's code and data in leased Virtual Machines, even when the powerful underlying hypervisor may be compromised. They use a non-bypassable form of hardware access control leveraging the hardware trend towards manycore chips and hardware virtualization features to enhance Cloud Security. They aim at exploring software-hardware co-design for security to design future trustworthy systems that provide security protections, at the levels needed, when needed, even when malware is in the system.

Another interesting work is by Cazalas et al. [8]. They study whether integrity of execution can be preserved for process-level virtualization protection schemes in the face of adversarial analysis. Their approach considers exploits that target the virtual execution environment itself and how it interacts with the underlying host operating system and hardware. Results indicate that such protection mechanisms may be vulnerable at the level where the virtualized code interacts with the underlying operating system, undermining security and calling for additional mitigation techniques using hardware-based integration or hybrid virtualization techniques that can better defend legitimate uses of virtualized software protection.

## 5 Secure Enclaves and Virtualization

In Cloud computing environments, hardware resources are shared, and parallel computation widespread that can produce privacy and security issues when isolation is not enforced. In fact, the hypervisor is an important cornerstone of Cloud computing that is not necessarily trustworthy or bug-free. To mitigate this threat Intel and AMD introduced respectively SGX [3] [9] and SEV [4] [29], which transparently encrypt a virtual machines memory. Intel introduced the SGX [11] hardware extensions to create a trusted execution environment (secure enclave or isolation container) within its CPUs. SGX claims runtime protection of a running process/VM even if the host OS and software components are malicious. Isolation containers are a primitive to minimize trusted software, leveraging trusted hardware and having a small performance overhead [11]. This is a smart idea though present implementations (AMD SEV and Intel SGX) do still have some limitations, as we detail in the following.

### 5.1 Intel SGX

Intel SGX [55] is an hardware technology aimed at protecting guest code and data from the hypervisor. It is an architecture extension designed to increase the security of software through an "inverse sandbox" mechanism. Legitimate software can be sealed inside an "enclave" and protected from unauthorized access, even when malware has hypervisor privileges. SGX was designed to comply with some clear requirements/objectives [9]:

– **protecting sensitive data** from unauthorized access or modification by rogue software running at higher privilege levels;
– **supporting legitimate software** allowing them to continue using platform resources;
– **maintaining consumer freedom** allowing them to retain control of their platforms and the freedom to install and uninstall applications and services as they choose;
– **allow certifying** an applications trusted code and produce a signed attestation, rooted in the processor, that includes this measurement and other certification that the code has been correctly initialized in a trustable environment;

---
[3] Software Guard Extensions
[4] Secure Encrypted Virtualization

- **supporting legacy** (development) tools, processes, and software distribution channels;
- **allowing scalability** of the performance of trusted applications in order to scale with the capabilities of the underlying hardware;
- **protecting applications** allowing them to define secure regions of code and data that maintain confidentiality even when an attacker has physical control of the platform and can conduct direct attacks on memory.

SGX minimizes the amount of code that provides support for the protected-module architecture, whereas module state persistence is delegated to the untrusted operating system. Nevertheless, state continuity must be guaranteed since an attacker should not be able to cause a module to use stale states (i.e. a rollback attack), and while the system is not under attack, a module should always be able to make progress, even when the system could crash or lose power at unexpected random points in time [55]. Providing state-continuity support is non-trivial as many algorithms are vulnerable to attack, require on-chip non-volatile memory, wear-out existing off-chip secure non-volatile memory and/or are too slow for many applications. ICE [55] is an interesting architecture providing state-continuity guarantees to protected modules by means of a machine-checked proof. ICE does not rely on secure non-volatile storage for every state update (e.g., the slow TPM chip) and is resilient to power losses.

### 5.2 SGX Security Issues

Albeit beneficial and promising in theory, the SGX approach has proven vulnerable to (mostly side-channel) attacks from its early days. As an example, CacheZoom [44] can track all memory accesses of SGX enclaves with high spatial and temporal precision. Further, AES key recovery attacks have been proven possible on SGX enclaves.

Hertzelt et al. [23] analyse to what extent the proposed features can resist a malicious hypervisor and discuss the tradeoffs imposed by additional protection mechanisms. They developed a model of SEV's security capabilities and found three design shortcomings. Firstly, the virtual machine control block is not encrypted and handled directly by the hypervisor, allowing it to bypass VM memory encryption by executing conveniently chosen gadgets. Secondly, the general purpose registers are not encrypted upon vmexit, leaking potentially sensitive data. Finally, the control over the nested pagetables allows a malicious hypervisor to closely monitor the execution state of a VM and attack it with memory replay attacks.

Schwarts et al [52] have found that SGX can be used to Conceal Cache Attacks. They demonstrate software-based side-channel attacks from a malicious SGX enclave targeting co-located enclaves, and abusing SGX protection features to conceal itself. The attack is fully functional even across multiple Docker containers. In fact the real issue with cache attacks lies with stealing information (such as private keys) rather that controlling a system.

Cloak [21] is another technique leveraging hardware transactional memory to prevent adversarial observation of cache misses on sensitive code and data. Cloak provides protection against cache-based side-channel attacks for SGX enclaves.

Constan's Sanctum [12] achieves stronger security guarantees under software attacks than SGX with an equivalent programming model. In fact, Sanctum offers the same promise as Intels Software Guard Extensions (SGX), namely strong provable isolation of software modules running concurrently and sharing resources, but protects against an important class of additional software attacks that infer private information from a programs memory access patterns. Sanctum reduces attack surface through isolation, rather than plugging attack-specific privacy leaks. Most of Sanctums logic is implemented in trusted software, which does not perform cryptographic operations using keys, and is easier to analyze than SGXs opaque microcode. Sanctum prototype leverages a RISC-V [58] core but is quite flexible in that it adds hardware at the interfaces between generic building blocks, replacing SGXs microcode with a software security monitor that runs at a higher privilege level than the hypervisor and the OS. On RISC-V, the security monitor runs at machine level, leveraging one privileged enclave, similarly to SGXs Quoting Enclave. The really interesting idea behind Sanctum is that it leverages a principled, transparent, and well-scrutinized approach to secure system design.

SGX may be vulnerable to other side channel attacks, such as cache access pattern monitoring (see also [5] by Brasser et al.). In fact, [5] proves that cache-based attacks are a serious threat to the confidentiality of SGX-protected programs by showing an attack without interrupting enclave execution. Brasser et al. also stress their approach has major technical challenges, since the existing cache monitoring techniques experience significant noise when the victim process is not interrupted.

The SGX-based branch shadowing attack is described in [33] which can reveal fine-grained control flows (i.e., each branch) of an enclave program running on real SGX hardware. In fact, SGX does not clear the branch history when switching from enclave mode to non-enclave mode, leaving the fine-grained traces to the outside world through a branch-prediction side channel. They developed two exploitation techniques: Intel PT- and LBR-based history-inferring techniques and APIC-based technique to control the execution of enclave programs in a fine-grained manner. As a result, their attack could brake ORAM, Sanctum, SGX-Shield, and T-SGX. A software-based countermeasure, called Zigzagger, was introduced by [33] to mitigate the branch shadowing attack in practice.

Brasser et al. [4] propose a data location randomization as a novel defensive approach against side-channel attacks. Their compiler-based tool called DR.SGX instruments enclave code to permute data locations at the granularity of cache lines. Brasser's solution protects most, but not all enclaves from typical SGX cache attacks.

## 6 Use Cases for Virtualization

This section introduces increasingly common Use Cases and Technological scenarios. One relevant topic is mobile virtualization for small devices such as smartphones, smart watches, and tablets, that are carried everywhere. They are referred to as Bring Your Own Device (BYOD) since their owner usually carries them even inside the secure perimeter of companies and, in general, at work. This section also highlights the usage of virtualization honeypots for malware collection and computer forensics purposes. In fact, malware can be analyzed and dissected based on the interaction with the emulated virtual environment.

### 6.1 BYOD and Virtualization

Personal mobile devices often enter enterprise boundaries. They can potentially hide malware or eavesdrop sensitive data to the outside world. At present, there is little or no control over an enterprise personnel mobile device data and application content and integrity. Banning such devices altogether from within enterprise boundaries does not seem a viable approach. A better one would imply remote attestation of the integrity and compliance of the employees mobile device to the desired security policies. Secure virtualization mechanisms based on a trusted transparent monitoring hypervisor would help. In fact, software integrity attestation future perspectives are good, given that ARM CPUs increasingly support virtualization extensions that allow implementing hypervisors that can run and monitor trusted VMs even on mobile/handheld devices [13]. The hypervisor would be able to enforce the exclusive execution of an enterprise VM when the device is inside well defined boundaries. The same VM can be disabled outside such boundaries in order to limit/prevent data breaches.

### 6.2 Virtualization and Smartphones

Increasingly often, smart mobile phones are relevant sources of information for investigations. Most currently available tools able to acquire forensic evidence from smartphones require destructive physical access to the device. This is one use case where secure virtualization can be used to access live data without interfering with regular phone activity and thus allowing live mobile forensics. LiveSD Forensics [6] is an example of on-device live data acquisition of the RAM and the EEPROM of Windows mobile devices. LiveSD Forensics uses a standard SD-card equipped with tailored code to perform the data acquisition. Unfortunately, LiveSD generates a memory alteration, albeit small.

In addition, virtualization allows creating mobile honeypots able to study and classify malware in a controlled way. In fact, similarly to mobile forensics, mobile virtualization can be used to collect malware and study its behavior, in a mostly transparent way. As mobile hardware is increasingly capable of running multiple VMs in parallel, different levels of security can be associated to different VMs to limit malware activity.

### 6.3 Future Research Directions

Future virtualization trends are mostly related to novel technological developments that aim at better isolation and performance. One such example is represented by ARM CPUs that, apart from being dominant in the mobile market, are increasingly present in the server arena. A second example is represented by Cloud-provided GPU access that is increasingly common. Finally, novel x86_64 processors integrate both CPU and GPU cores. Nevertheless, they have to provide additional security guarantees. Efficiently virtualizing distributed heterogeneous computing in the Cloud is an opportunity to improve

Cloud security and reliability. Further, in order to allow efficient secure usage of multicores, such resources have to be constantly monitored for anomalous usage patterns, since sharing resources also introduces additional security and privacy issues. Finally, the availability of an increasingly large amount of computing cores allows using them for a number of novel applications, such as computation replication for reliability and availability or proactive computing for most different possible scenarios.

## 7 Conclusion

Virtualization is at the heart of Cloud computing. Albeit more lightweight approaches such as Containerization and Unikernels exist, hardware-supported isolation mechanisms provide beneficial in many different scenarios where security requirements are relevant. Nevertheless, security vulnerabilities are still a major issue, as highlighted by recently discovered exploits. Enhanced virtualization approaches and more effective isolation and monitoring technologies, that can also leverage additional computing resources of recent CPUs and GPUs, are still in their infancy. Such advances, coupled with appropriate software counterparts, will possibly improve the integrity and security of resources in Cloud, server farms, and in mobile scenarios.

Virtualization Technologies and Cloud Security    179. Chakrabarti, S., Leslie-Hurd, R., Vij, M., McKeen, F., Rozas, C., Caspi, D., Alexandrovich, I., Anati, I.: Intel software guard extensions (intel; sgx) architecture for oversubscription of secure memory in a virtualized environment. In: Proc Hardware and Architectural Support for Security and Privacy. pp. 7:1–7:8. HASP '17, ACM, New York, NY, USA (2017)
10. Combe, T., Martin, A., Di Pietro, R.: To docker or not to docker: A security perspective. IEEE Cloud Computing 3(5), 54–62 (2016)
11. Costan, V., Lebedev, I., Devadas, S.: Secure processors part i: Background, taxonomy for secure enclaves and intel sgx architecture. Foundations and Trends in Electronic Design Automation 11(1-2), 1–248 (2017)
12. Costan, V., Lebedev, I.A., Devadas, S.: Sanctum: Minimal hardware extensions for strong software isolation. In: USENIX Security Symp. pp. 857–874 (2016)
13. Dall, C., Nieh, J.: Kvm/arm: The design and implementation of the linux arm hypervisor. SIGARCH Comput. Archit. News 42(1), 333–348 (Feb 2014)
14. Di Pietro, R., Franzoni, F., Lombardi, F.: HyBIS: Advanced introspection for effective windows guest protection. In: ICT Systems Security and Privacy Protection. pp. 189–204. Springer Intl. Publishing (2017)
15. Di Pietro, R., Lombardi, F., Signorini, M.: CloRExPa: Cloud Resilience via Execution Path Analysis. Future Gener. Comput. Syst. 32, 168–179 (mar 2014)
16. Di Pietro, R., Lombardi, F., Villani, A.: CUDA Leaks: A detailed hack for cuda and a (partial) fix. ACM Trans. Embed. Comput. Syst. 15(1), 15:1–15:25 (Jan 2016)
17. Dowty, M., Sugerman, J.: GPU virtualization on vmware's hosted i/o architecture. SIGOPS Oper. Syst. Rev. 43(3), 73–82 (Jul 2009)
18. Dua, R., Raja, A.R., Kakadia, D.: Virtualization vs containerization to support paas. In: 2014 IEEE Intl. Conf. on Cloud Engineering. pp. 610–614 (March 2014)
19. By Hertzsprung at English Wikipedia, C.B.S..: Execution rings. https://commons.wikimedia.org/w/index.php?curid=8950144
20. Fernandes, D.A.B., Soares, L.F.B., Freire, M.M., Incio, P.R.M.: Randomness in virtual machines. In: 2013 IEEE/ACM 6th Intl. Conf. on Utility and Cloud Computing. pp. 282–286 (Dec 2013)
21. Gruss, D., Lettner, J., Schuster, F., Ohrimenko, O., Haller, I., Costa, M.: Strong and efficient cache side-channel protection using hardware transactional memory. In: 26th USENIX Security Symp. (USENIX Security 17). pp. 217–233. USENIX Association, Vancouver, BC (2017)
22. Gupta, V., Gavrilovska, A., Schwan, K., Kharche, H., Tolia, N., Talwar, V., Ranganathan, P.: GViM: GPU-accelerated virtual machines. In: Proc. of the 3rd ACM Workshop on System-level Virtualization for High Performance Computing. pp. 17–24. HPCVirt '09, ACM, New York, NY, USA (2009)
23. Hetzelt, F., Buhren, R.: Security analysis of encrypted virtual machines. SIGPLAN Not. 52(7), 129–142 (Apr 2017)
24. Hong, C.H., Spence, I., Nikolopoulos, D.S.: Gpu virtualization and scheduling methods: A comprehensive survey. ACM Comput. Surv. 50(3), 35:1–35:37 (Jun 2017)
25. Intel: Intel virtualization technology specification for the ia-32 intel architecture. http://dforeman.cs.binghamton.edu/~foreman/550pages/Readings/intel05virtualization.pdf last accessed 2018-02-02 (2005)
26. Jason, G.: VENOM: Virtualized Environment Neglected Operations Manipulation. Available from MITRE, CVE-ID CVE-2015-3456. (May 2015)
27. Jia, L., Zhu, M., Tu, B.: T-vmi: Trusted virtual machine introspection in cloud environments. In: Proc. of the 17th IEEE/ACM Intl. Symp. on Cluster, Cloud and Grid Computing. pp. 478–487. CCGrid '17, IEEE Press, Piscataway, NJ, USA (2017)